\documentclass[pra,twocolumn,showpacs,preprintnumbers,amsmath,amssymb]{revtex4-1}

\usepackage{graphicx}
\usepackage{dcolumn}
\usepackage{bm}
\usepackage{color}

\usepackage{enumerate}

\def\rr{{\bm r}}
\def\kk{{\bm k}}
\def\vac{|{\rm vac}\rangle}
\newcommand{\dha}[1]{\hat{a}^\dagger_{#1}}
\newcommand{\ha}[1]{\hat{a}_{#1}}
\newcommand{\dhaa}[1]{\hat{\alpha}^\dagger_{#1}}
\newcommand{\haa}[1]{\hat{\alpha}_{#1}}
\newcommand{\ketPsi}[1]{\left|\Psi{#1}\right\rangle}
\newcommand{\braPsi}[1]{\left\langle\Psi{#1}\right|}
\newcommand\non\nonumber

\begin{document}

\preprint{APS/123-QED}


\title{Goldstone-mode Instability leading to Fragmentation in a Spinor Bose-Einstein Condensate}
\author{Yuki Kawaguchi}
\affiliation{
Department of Applied Physics and Quantum-Phase Electronics Center, University of Tokyo, 7-3-1 Hongo, Bunkyo-ku, Tokyo 113-0032, Japan
}
\date{\today}

\begin{abstract}
We apply the number-conserving Bogoliubov theory to spinor Bose-Einstein condensates and show that the Goldstone magnon leads instability leading to fragmentation.
Unlike the dynamical instability, where modes with complex eigenfrequencies grow exponentially, here the zero-energy mode exhibits algebraic growth.
We also point out that a small fraction of thermally excited atoms enhances the fragmentation dynamics.
\end{abstract}


\maketitle

\section{introduction}
When a symmetry is spontaneously broken in an ordered state, the excitation that recovers the broken symmetry is gapless.
This is the well-known Goldstone theorem~\cite{Goldstone}.
Typical examples include the phonon in a Bose-Einstein condensate (BEC) and the magnon in a Heisenberg ferromagnet, where the U(1) gauge symmetry and the SO(3) spin rotational symmetry are broken, respectively.

When the system is finite, however, the symmetry breaking is not always exact.
For example, the phase of the order parameter diffuses in a trapped BEC, and the U(1) symmetry is recovered in time evolution~\cite{Lewenstein1996}.
Another example is a system of spin-1 atoms with antiferromagnetic interactions:
When the atoms are confined in a tiny trap so that the motional degrees of freedom are frozen, the exact ground state is a BEC of spin-singlet pairs, which preserves the SO(3) spin rotational symmetry~\cite{Law1998,Koashi2000}.
On the other hand, the mean-field theory assumes that all atoms are condensed in a single-particle state, which always breaks the SO(3) symmetry.
Hence, the mean-field ground state for antiferromagnetic interactions, which is called the polar state~\cite{Ho1998}, is unstable in a micro condensate~\cite{Law1998,Pu1999,Zhou2001}.
However, since the difference in the energy per atom between the mean-field and the exact ground states is proportional to $1/N$, where $N$ is the total number of atoms, an infinitesimal fluctuation breaks the symmetry of the system and stabilizes the mean-field state in the thermodynamic limit.

In this paper, we examine the stability of Goldstone magnons in spinor BECs, and show that they become unstable in a finite system, explaining the instability of the polar state.
The time scale of the instability diverges in the thermodynamic limit, and therefore the mean-field state becomes stable.
This instability leads to fragmentation of the condensate~\cite{Nozieres1982,Mueller2006}.
According to the Penrose-Onsager criterion~\cite{PenroseOnsager}, a system is Bose-Einstein condensed when the single-particle density matrix has an eigenvalue of order $N$.
If there is only one eigenvalue that is of order $N$, the atoms are condensed in a single-particle state, whose wave function corresponds to the order parameter in the mean-field theory.
On the other hand, in a fragmented BEC, there are several eigenvalues of order $N$.
For the case of a polar BEC, two eigenvalues of the single-particle density matrix increase in time.
The quadratic Zeeman energy dependence of this instability has been investigated theoretically~\cite{Cui2008,Barnett2010,*Barnett2011} and experimentally~\cite{Bookjans2011,Vinit2013}.

The instabilities in BECs, so far, have been mainly discussed in the context of the dynamical instability, which is characterized by a complex eigenvalue of the Bogoliubov equation.
In contrast to the Landau instability, which is a negative-energy excitation and grows as the energy dissipates, the dynamical instability grows exponentially even in the absence of energy dissipation, explaining many phenomena in a trapped BEC, such as dumping of the superfluid flow in an optical lattie~\cite{Wu2001} and quench dynamics in spin-1~\cite{Sadler2006,*Saito2007,*Lamacraft2007,*Uhlmann2007,*Sau2009} and spin-2~\cite{Klempt2009,*Klempt2010,*Scherer2010} BECs.
As for zero-energy modes, it is shown that there is a nonzero contribution of Goldstone phonons in the Bogoliubov Hamiltonian, which causes a diffusion of the condensate phase in a scalar BEC~\cite{Lewenstein1996} and that of the spin direction in a ferromagnetic BEC~\cite{Yi2003a,*Yi2003b}.
Here, we apply the number-conserving Bogoliubov theory~\cite{Leggett2001}, which does not assume the U(1) symmetry breaking, to spinor BECs~\cite{Kawaguchi2012,Stamper-Kurn2013},
and find that the Goldstone magnon exhibits algebraic growth, leading to fragmentation.
In contrast to the instability of Goldstone phonons whose wave function is the same as the condensate, the wave function for the Goldstone magnon is orthogonal to the condensate.
Hence, the atoms in the magnon mode are distinct from the condensed atoms, and the initial amount of the Goldstone magnons is tunable in experiments.
This tunability enables us to control fragmentation dynamics, because the growth of the Goldstone magnon is enhanced due to bosonic stimulation when the magnon mode is initially occupied.

The rest of this paper is organized as follows.
In Sec.~II, we review the system of a spin-1 polar BEC.
In Sec.~III, we apply the number-conserving Bogoliubov theory to spinor BECs:
We introduce the variational wave function for a fixed number state,
and derive the equation of motion for the variational parameters.
In Sec.~IV, by solving the equation of motion, we show that the Goldstone magnons exhibit algebraic growth.
The growth is enhanced by initially populated atoms in the zero mode, which is confirmed by numerical simulation.
A possible experimental scheme for observing the fragmentation dynamics is also discussed in Sec.~IV.
Section~V concludes this paper.

\section{Spin-1 Polar BEC}
The Hamiltonian of a spin-1 BEC in a uniform system is given by~\cite{Ohmi1998,Ho1998} 
\begin{align}
 \hat{H} = \sum_{m,\kk}(\epsilon_{\kk} +q m^2) \dha{m \kk} \ha{m \kk} + \hat{V},
\end{align}
where $\hat{a}_{m\kk}$ annihilates an atom with momentum $\kk$ in magnetic sublevel $m_F=m\,(=0,\pm1)$, $\epsilon_k\equiv \hbar^2\kk^2/(2M)$ with $M$ being the atomic mass, and $q$ is the quadratic Zeeman energy per atom.
The summation with respect to $m$ is taken for $m=0,\pm1$ unless otherwise noted.
For a spin-1 system, the interaction part is divided into the density-density interaction and the spin-exchange interaction and is given by
\begin{align}
\hat{V} =
& \frac{1}{2\Omega}\sum_{m_1m_2m_3m_4}\sum_{\kk_1\kk_2\kk_3\kk_4} \non\\
&
\left[c_0\delta_{m_1m_4}\delta_{m_2m_3} + c_1(\bm F)_{m_1m_4}\cdot (\bm F)_{m_2m_3}\right]
\non\\
&
\delta_{\kk_1+\kk_2,\kk_3+\kk_4}
\dha{m_1 \kk_1}\dha{m_2 \kk_2}\ha{m_3 \kk_3}\ha{m_4 \kk_4},
\label{eq:Hamiltonian}
\end{align}
where $\Omega$ is the volume of the system, $\bm F$ is the vector of the spin-1 spin matrices, and the interaction coefficients are given by $c_0=4\pi\hbar^2 (2a_2+a_0)/(3M)$ and $c_1=4\pi\hbar^2 (a_2-a_0)/(3M)$ with $a_S$ being the scattering length of two colliding atoms with total spin $S$.
The interaction Hamiltonian is also written as
\begin{align}
 \hat{V} &= \frac{1}{2\Omega}\sum_{\kk_1\kk_2\kk_3\kk_4}\delta_{\kk_1+\kk_2,\kk_3+\kk_4} \non\\
&\bigg[
(c_0+c_1)\sum_{m_1m_2}\dha{m_1 \kk_1}\dha{m_2 \kk_2}\ha{m_2 \kk_3}\ha{m_1 \kk_4}
\label{eq:Vint2}\\
&-c_1 (2\dha{1 \kk_1}\dha{-1 \kk_2} - \dha{0 \kk_1}\dha{0 \kk_2})(2\ha{1 \kk_3}\ha{-1 \kk_4} - \ha{0 \kk_3}\ha{0 \kk_4})\bigg].
\non
\end{align}
In the following, we consider an antiferromagnetic interaction, i.e., $c_1>0$.

We investigate the instability of the mean-field polar state in which all atoms are condensed in the $m_F=0$ state.
The corresponding $N$-particle state is given by
\begin{align}
\ketPsi{_\textrm{MF}}&= \frac{1}{\sqrt{N!}}\left(\dha{0 \bm0}\right)^N\vac,
\label{eq:PsiMF}
\end{align}
where $\vac$ denotes the vacuum of atoms.
$\ketPsi{_\textrm{MF}}$ is the mean-field ground state for $q>0$. 
For a negative $q$, the mean-field ground state is the superposition of the $m_F=1$ and $-1$ states, which is related to $\ketPsi{_\textrm{MF}}$ by a $\pi/2$ rotation about an axis perpendicular to the $z$ axis.
At $q=0$, these states are degenerate, because the Hamiltonian is invariant under SO(3) rotations in the spin space.

The Bogoliubov spectra for the polar state are composed of one phonon branch and two degenerate magnon branches, whose dispersions are respectively given by~\cite{Ohmi1998,Kawaguchi2012}
\begin{align}
E_\kk^\textrm{ph}&=\sqrt{\epsilon_\kk(\epsilon_\kk+2c_0 n)},
\label{eq:phonon}\\
E_\kk^\textrm{mag}&=\sqrt{(\epsilon_\kk+q)(\epsilon_\kk+q+2c_1 n)},
\label{eq:magnon}
\end{align}
where $n\equiv N/\Omega$.
At $q=0$ the magnon modes become gapless, corresponding to the Goldstone modes associated with the SO(3) spin rotation symmetry.
When $q<0$, $E_\kk^\textrm{mag}$ becomes purely imaginary for long wavelengths, indicating that the polar state is dynamically unstable.

\section{Number-conserving Bogoliubov Theory}
We consider dynamics starting from $\ketPsi{_\textrm{MF}}$.
Since $|\Psi_{\rm MF}\rangle$ is not the exact ground state, it evolves in time.
Here, we diagnose the stability of the polar state
by considering whether the number of atoms in each $(m,\kk)\neq(0,\bm0)$ state remains much smaller than $N$.
Since the interaction terms in the Hamiltonian~\eqref{eq:Hamiltonian} conserve the total spin and the total momentum of two colliding atoms, pairs of atoms in magnetic sublevels $m$ and $-m$ with momentum $\kk$ and $-\kk$ are created from $\ketPsi{_\textrm{MF}}$ via pairing terms $\dha{0\kk}\dha{0,-\kk}\ha{0 \bm0}\ha{0 \bm0}$ and $\dha{1\kk}\dha{-1,-\kk}\ha{0 \bm0}\ha{0 \bm0}$ [see Eq.~\eqref{eq:Vint2}].
We hence introduce the following variational wave function for the $N$-particle state:
\begin{align}
\ketPsi{(t)}
=& \frac{\mathcal{N}}{\sqrt{N!}} \bigg[ (\dha{0\bm0})^2
- \sum_{\kk\neq\bm 0} \Lambda_{\kk}(t) \dha{0\kk}\dha{0,-\kk} \non\\
&-\sum_{\kk}\Lambda'_{\kk}(t)\dha{1\kk}\dha{-1,-\kk}\bigg]^{N/2}\vac,
\label{eq:Psi1}
\end{align}
where $\mathcal{N}$ is the normalization coefficient, $\Lambda_\kk$ and $\Lambda_\kk'$ satisfy $\Lambda_{-\kk}=\Lambda_\kk$ and $\Lambda'_{-\kk}=\Lambda'_\kk$, and we assume that $N$ is even for simplicity.
In contrast to the mean-field (Hartree) approximation, in which all atoms are condensed in the same single-particle state, the above ansatz assumes that all pairs of atoms occupy the same two-particle state.
In this sense, Eq.~\eqref{eq:Psi1} is a natural expansion of the Hartree approximation so as to include two-particle correlations,
and hence it is adequate for discussing physics related to two-particle correlations.
Although we need a more general wave function that includes three-particle or higher correlations to describe the exact dynamics,
our present interest is not the exact dynamics but how the instability develops starting from the mean-field state.
Since the elementary process of this instability is the two-body scattering of condensed atoms into excited states, two-particle correlation is enough to predict the instability.

For the sake of systematic formulation, we define $\haa{\pm1,\kk}\equiv (\ha{1\kk}\pm\ha{-1\kk})/\sqrt{2}$ and $\haa{0 \kk}\equiv \ha{0 \kk}$ which satisfy the bosonic commutation relation: $[\haa{m\kk},\dhaa{m'\kk'}] = \delta_{mm'}\delta_{\kk,\kk'}$.
Then the variational ansatz~\eqref{eq:Psi1} is rewritten as
\begin{align}
\ketPsi{(t)}
=& \frac{\mathcal{N}}{\sqrt{N!}} \bigg[ (\dhaa{0\bm0})^2
- \sum_{m=\pm1} \lambda_{m\bm0}(t) (\dhaa{m\bm0})^2 \non\\
&-\sum_{m,\kk\neq0}\lambda_{m\kk}(t)\dhaa{m\kk}\dhaa{m,-\kk}\bigg]^{N/2}\vac,
\label{eq:Psit}
\end{align}
where $\lambda_{0\kk}=\Lambda_\kk$, $\lambda_{1\kk}=-\lambda_{-1\kk}=\Lambda'_\kk/2$, and $\lambda_{m,-\kk}=\lambda_{m\kk}$.
The Hamiltonian in the basis of $\alpha$-particles is given by
\begin{align}
 \hat{H} =& \sum_{m\kk}(\epsilon_{\kk} +q m^2) \dhaa{m \kk} \haa{m \kk} \non\\
+& \frac{1}{2\Omega}\sum_{\kk_1\kk_2\kk_3\kk_4}\delta_{\kk_1+\kk_2,\kk_3+\kk_4} \non\\
&\bigg[
(c_0+c_1)\sum_{m_1m_2}\dhaa{m_1 \kk_1}\dhaa{m_2 \kk_2}\haa{m_2 \kk_3}\haa{m_1 \kk_4} 
\label{eq:H_alpha}\\
&-c_1 (\dhaa{1 \kk_1}\dhaa{1 \kk_2} -\dhaa{0 \kk_1}\dhaa{0 \kk_2} - \dhaa{-1, \kk_1}\dhaa{-1, \kk_2})\non\\
&\ \ \ \ \ \ \ \ \times(\haa{1 \kk_3}\haa{1 \kk_4} - \haa{0 \kk_3}\haa{0 \kk_4} - \haa{-1, \kk_3}\haa{-1 \kk_4})\bigg].
\non
\end{align}

Following the framework of the Lagrangian formulation, the functional action of the time-dependent Schr\"odinger equation is given by
\begin{align}
S[\{\lambda_{m\kk},\lambda^*_{m\kk}\}] =\int dt \braPsi{(t)} \hat{H}-i\hbar\frac{d}{d t}\ketPsi{(t)}.
\label{eq:action_def}
\end{align}
The equation of motion of $\lambda_{m\kk}(t)$ is obtained by taking the variation of the action $S$ with respect to the coefficients $\{\lambda_{m\kk}\}$:
\begin{align}
 \frac{\delta S[\{\lambda_{m\kk},\lambda^*_{m\kk}\}]}{\delta \lambda^*_{m\kk}}=0.
\label{eq:action}
\end{align}
In the following calculation, we assume that the total depletion of the condensate is much smaller than $N$, which requires $|\lambda_{m\kk}|<1$ for all $(m,\kk)\neq(0,\bm0)$ [see Eq.~\eqref{eq:aa}].
This assumption allows us to neglect the terms in the order of $N^{-1}$ in evaluating the action, as in the case of the conventional Bogoliubov theory.
As we will see in the next section, for some cases $|\lambda_{m\kk}|$ goes to unity as $t\to \infty$, and the fraction of the corresponding mode diverges.
In such cases, the polar state is unstable and the ansatz~\eqref{eq:Psit} eventually becomes invalid.
We are interested in how such an instability grows at the beginning.

To calculate the normalization coefficient $\mathcal{N}$, we use the multinomial expansion
\begin{align}
&\left[(\dhaa{0\bm0})^2-\sum_{m=\pm1} \lambda_{m\bm0}(\dhaa{m\bm0})^2 - \sum_{m,\kk>0} 2\lambda_{m\kk}\dhaa{m\kk}\dhaa{m,-\kk}\right]^{N/2}\non\\
&= \left(\frac{N}{2}\right)! \sum_{\{p_{m\kk}\}} 
\frac{(\dhaa{0\bm0})^{N-2N'}}{(N/2-N')!}
\prod_{m'=\pm1}\frac{[-\lambda_{m'\bm0}(\dhaa{m'\bm0})^2]^{p_{m'\bm0}}}{p_{m'\bm0}!}\non\\
&\hspace{5mm}\times
\prod_{m,\kk>0} \frac{(-2\lambda_{m\kk}\dhaa{m\kk}\dhaa{m,-\kk})^{p_{m\kk}}}{p_{m\kk}!}
\end{align}
where $N'\equiv \sum_{m=\pm1}p_{m\bm0}+\sum_{m,\kk>0}p_{m\kk}$, the $p_{m\kk}$'s are non-negative integers satisfying $N'\le N/2$, 
$\sum_{\{p_{m\kk}\}}$ means summation for all possible combinations of $\{p_{m\kk}\}$, and $\sum_{\kk>0}$ and $\prod_{\kk>0}$ count the contribution of either $\kk$ or $-\kk$.
Then $\mathcal{N}$ is calculated as
\begin{align}
\mathcal{N}^{-2} 
=& \sum_{\{p_{m\kk}\}} \frac{(N-2N')!}{N!} \left[\frac{(N/2)!}{(N/2-N')!}\right]^2
\non\\
&\times \prod_{m'=\pm1}\frac{|\lambda_{m\bm0}|^{2p_{m\bm0}}(2p_{m\bm0})!}{(p_{m\bm0}!)^2} 
 \prod_{m,\kk>0}|2\lambda_{m\kk}|^{2p_{m\kk}}\non\\
=& \sum_{\{p_{m\kk}\}}   \frac{g(N/2-N')  }{g(N/2)}\non\\
 &\times  \prod_{m'=\pm1}\prod_{m,\kk>0} g(p_{m'\bm0}) |\lambda_{m'\bm0}|^{2p_{m'\bm0}}  |\lambda_{m\kk}|^{2p_{m\kk}},\label{eq:norm1}
\end{align}
where $g(x)=4^{-x}\Gamma(2x+1)/[\Gamma(x+1)]^2$, with $\Gamma(x)\equiv \int_0^\infty t^{x-1}e^{-t}dt$ being the Gamma function, is a monotonically decreasing function satisfying $g(0)=1$ and $g(x)\to 1/\sqrt{\pi x}$ as $x\to \infty$.
Note that $g(N/2-N')/g(N/2)$ is monotonically increasing as a function of $N'$ and satisfies $g(N/2-N')/g(N/2)\le \sqrt{\pi N'}$ for $0\le N'\le N/2$. 
In addition, it can be approximated as $g(N/2-N')/g(N/2)\sim 1+N'/N$ for $N'\ll N/2$. 
On the other hand, because $|\lambda_{m\kk}|<1$ by assumption, 
the last line of Eq.~\eqref{eq:norm1} decays faster than exponentially: $\prod_{m'=\pm1}\prod_{m,\kk>0} g(p_{m'\bm0}) |\lambda_{m'\bm0}|^{2p_{m'\bm0}}  |\lambda_{m\kk}|^{2p_{m\kk}} \le \delta^{N'}$ where $\delta\equiv \max_{(m,\kk)\neq (0,\bm0)}|\lambda_{m\kk}|^2<1$.
Hence, the main contribution of Eq.~\eqref{eq:norm1} comes from the region of $N' \ll N/2$.
We then approximate $g(N/2-N')/g(N/2)\sim 1$ and remove the constraint $\sum_{m=\pm1}p_{m,\bm0}+\sum_{m,\kk>0}p_{m\kk}\le N/2$ in taking the summation $\sum_{\{p_{m\kk}\}}$, obtaining
\begin{align}
\mathcal{N}^2 = \prod_{m'=\pm1} \sqrt{1-|\lambda_{m'\bm0}|^2}\prod_{m,\kk>0}  (1-&|\lambda_{m\kk}|^2) + O(1/N),
\label{eq:mathcalN}
\end{align}
where we have used the formulas
$\sum_{p=0}^{\infty} g(p) x^{2p} =1/(1-x^2)^{1/2}$ and $\sum_{p=0}^{\infty} x^{2p} = 1/(1-x^2)$.
In a similar manner, we obtain 
\begin{align}
\langle \dhaa{m\kk}\haa{m\kk} \rangle &= \frac{|\lambda_{m\kk}|^2}{1-|\lambda_{m\kk}|^2} + O(1/N),
\label{eq:aa}\\
\langle \dhaa{0\bm0}\haa{0\bm0}\dhaa{m\kk}\haa{m\kk} \rangle &= N \left[\frac{|\lambda_{m\kk}|^2}{1-|\lambda_{m\kk}|^2} + O(1/N)\right],\\
\langle \dhaa{m\kk}\dhaa{m,-\kk}\haa{0\bm0}\haa{0\bm0} \rangle &= - N \left[\frac{\lambda_{m\kk}^*}{1-|\lambda_{m\kk}|^2} + O(1/N)\right],
\end{align}
for $(m,\kk)\neq(0,\bm0)$, where $\langle\cdots\rangle\equiv \braPsi{(t)}\cdots\ketPsi{(t)}$ and we have used $ \sum_{p=0}^{\infty} 2p g(p) x^{2p} = x^2/(1-x^2)^{3/2}$ and $ \sum_{p=0}^{\infty} p x^{2p} = x^2/(1-x^2)^{2}$.
Because the total number of atoms is conserved, the number of condensed atoms can be rewritten as $\dhaa{0\bm0}\haa{0\bm0}=N-\sum_{(m,\kk)\neq (0,\bm0)}\dhaa{m\kk}\haa{m\kk}$, and we have
\begin{align}
 \langle \dhaa{0\bm0}\dhaa{0\bm0}\haa{0\bm0}\haa{0\bm0}\rangle 
&= N\left[N-1 +2\frac{|\lambda_{m\kk}|^2}{1-|\lambda_{m\kk}|^2}
+O(1/N)\right].
\label{eq:aaaa}
\end{align}
Here, we used the fact that $\langle \dhaa{m\kk}\haa{m\kk}\rangle$ is of order 1 from the assumption and hence $\langle \dhaa{m\kk}\haa{m\kk}\dhaa{m'\kk'}\haa{m'\kk'}\rangle=O(1)$ for $(m,\kk)\neq(0,\bm0)$ and $(m',\kk')\neq (0,\bm0)$.
Using the above results, the expectation value for the Hamiltonian~\eqref{eq:H_alpha} is given by
\begin{align}
 \langle \hat{H} \rangle
=& \frac{c_0n}{2}(N-1) \non\\
&+ \sum_{(m,\kk)\neq (0,\bm0)}
\frac{(\epsilon_\kk + A_m)|\lambda_{m\kk}|^2 - B_m {\rm Re}(\lambda_{m\kk})}{1-|\lambda_{m\kk}|^2}\non\\
&+O(1/N),
\label{eq:<H>}
\end{align}
where $A_{\pm1}=q+c_1n, A_{0}=c_0n, B_{\pm1}=\pm c_1n$, and $B_0=c_0n$.
Note that the terms of order $1/N$ in Eqs.~\eqref{eq:mathcalN}--\eqref{eq:aaaa}
depend on the $\lambda_{m\kk}$'s
and their contribution to $\langle \hat{H} \rangle$ is of order $1/N$ even after taking the summation with respect to $m$ and $\kk$.

The time derivative of the variational wave function is calculated as follows:
\begin{align}
&\langle \Psi(t)| \frac{d}{d t}|\Psi(t)\rangle \non\\
&= \frac{\dot{\mathcal{N}}}{\mathcal{N}}
 - \frac{\mathcal{N}^2}{N!} \langle \textrm{vac}| \left[\haa{0}^2- \cdots \right]^{N/2} \non\\
&\hspace{10mm}\frac{N}{2}[\sum_{m=\pm1}\dot{\lambda}_{m\bm0}\dhaa{m\bm0}\dhaa{m\bm0}+\sum_{m,\kk>0}2\dot{\lambda}_{m\kk}\dhaa{m\kk}\dhaa{m-\kk}]\non\\
&\hspace{30mm}\left[(\dhaa{0})^2-\cdots\right]^{N/2-1}\vac \non\\
&=
\sum_{(m,\kk)\neq(0,\bm0)}\frac{\lambda_{m\kk}^*\dot{\lambda}_{m\kk}-\dot{\lambda}_{m\kk}^* \lambda_{m\kk}}{4(1-|\lambda_{m\kk}|^2)}
+O(1/N),
\label{eq:dPdt}
\end{align}
where $\dot{x}\equiv dx/dt$.
Substituting Eqs.~ \eqref{eq:action_def}, \eqref{eq:<H>} and \eqref{eq:dPdt} into Eq.~\eqref{eq:action}
and neglecting terms of order $1/N$ in the action,
we finally obtain the equation of motion for $\lambda_{m\kk}$:
\begin{align}
-i\hbar \dot{\lambda}_{m\kk} &= B_m (\lambda_{m\kk}^2 + 1) - 2 (\epsilon_\kk + A_{m}) \lambda_{m\kk}.
\label{eq:dcdt}
\end{align}

\section{Growth of Goldstone Magnons}
We first consider the dynamics starting from $\lambda_{m\kk}(t=0)=0$.
In this case, the solution of Eq.~\eqref{eq:dcdt} is given by
\begin{align}
&\lambda_{m\kk}(t) = \frac{B_m\sin(E_{m\kk}t/\hbar)}{(\epsilon_\kk+A_{m\kk})\sin(E_{m\kk}t/\hbar)-iE_{m\kk}\cos(E_{m\kk}t/\hbar)},
\end{align}
where
\begin{align}
E_{m\kk}=\sqrt{(\epsilon_\kk+A_m+B_m)(\epsilon_\kk+A_m-B_m)}
\end{align}
reproduces the phonon spectrum [Eq.~\eqref{eq:phonon}] for $m=0$ and the magnon spectrum [Eq.~\eqref{eq:magnon}] for $m=\pm1$.
The number of excited atoms evolves as
\begin{align}
N_{m\kk}(t)\equiv\langle \dhaa{m\kk}\haa{m\kk} \rangle = \left|\frac{B_m}{E_{m\kk}}\sin\left(\frac{E_{m\kk}t}{\hbar}\right)\right|^2.
\label{eq:Nmk}
\end{align}
When the $E_{m\kk}$'s are real for all $(m,\kk)\neq (0,\bm 0)$, 
each $N_{m\kk}$ oscillates with frequency $2E_{m\kk}/\hbar$ due to quantum fluctuation.
Because all the $N_{m\kk}$'s are finite, the initial polar state is stable.
The polar state becomes unstable when one or some of the $N_{m\kk}$'s diverge.
Since $\langle \dhaa{m\kk}\haa{m'\kk}\rangle=0$ for $m\neq m'$, the increase in $N_{m\kk}$ means a fragmentation of the condensate.
Clearly, the polar state becomes unstable when there is an imaginary $E_{m\kk}$.
This is the case of the dynamical instability where the unstable modes grow exponentially~\cite{Wu2001,Sadler2006,*Saito2007,*Lamacraft2007,*Uhlmann2007,*Sau2009,Klempt2009,*Klempt2010,*Scherer2010}.
The present formalism also reproduces the exponential growth:
As time evolves, $|\lambda_{m\kk}|$ goes to unity and the number of the corresponding mode exponentially increases as $N_{m\kk}\sim e^{2|E_{m\kk}|t/\hbar}$.
Starting from the polar state, the dynamical instability occurs for $q<0$ and 
the number of atoms in the $m_F=\pm1$ states exponentially increases.

The system also becomes unstable when there is a zero-energy mode, because Eq.~\eqref{eq:Nmk} at $E_{m\kk}=0$ reduces to
\begin{align}
N_{m\kk}(t) = \left(\frac{B_m t}{\hbar}\right)^2.
\label{eq:Nmk0}
\end{align}
Unlike the exponential growth of the dynamical instability, the zero-energy mode grows in proportion to $t^2$.
For the polar state at $q=0$, the Goldstone magnons cause this instability, and the fraction of the Goldstone magnons ($N_{\pm1,\bm 0}/N$) increases with the characteristic time scale of $\tau_0=\hbar\sqrt{N}/(c_1n)$, which diverges in the thermodynamic limit.

Next, we consider the case when a small fraction of atoms is initially excited.
We solve Eq.~\eqref{eq:dcdt} with the initial condition $\lambda_{m\kk}(0)=\sqrt{N_{\rm ex}/(N_{\rm ex}+1)}e^{i\theta}$, which satisfies $N_{m\kk}(0)=N_{\rm ex}$.
Here, $\lambda_{m\kk}(0)$ includes an arbitrary phase $\theta$, which affects subsequent dynamics.
When the excited atoms do not have coherence with the condensed atoms, however, it is natural to take an average over all possible values of $\theta$, resulting in
\begin{align}
\bar{N}_{m\kk}(t) &= \frac{1}{2\pi}\int_0^{2\pi} d\theta N_{m\kk}(t) \nonumber\\
&= N_{\rm ex} + (2N_{\rm ex}+1) \left|\frac{B_m}{E_{m\kk}}\sin\left(\frac{E_{m\kk}t}{\hbar}\right)\right|^2.
\label{eq:barN}
\end{align}
Compared with Eq.~\eqref{eq:Nmk}, the time-dependent part of $\bar{N}_{m\kk}(t)$ is enhanced by a factor of $2N_{\rm ex}+1$ due to bosonic stimulation.
This enhancement is significant for the algebraic growth of zero modes because the time scale for the growth is shortened by a factor of $1/\sqrt{2N_{\rm ex}+1}$.
On the other hand, the time scale for the exponential growth is hardly affected when $|E_{m\kk}|$ is large enough.

\begin{figure}[htb]
 \includegraphics[width=\linewidth]{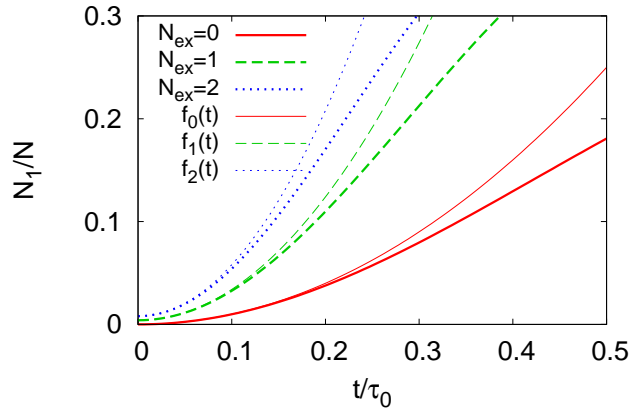}
 \caption{(Color online) Time evolution of $N_1/N$ at $q=0$ starting from the Fock state $|N_{\rm ex},N_{\rm ex}\rangle$.
The thick curves are numerically calculated following the Hamiltonian~\eqref{eq:Hsimple} for $N=250$ and $N_{\rm ex}=0$ (solid curve), 1 (dashed curve), and 2 (dotted curve).
The thin curves show the prediction based on Eq.~\eqref{eq:barN} where $f_{N_{\rm ex}}(t)=N_{\rm ex}/N+\sqrt{2N_{\rm ex}+1}(t/\tau_0)^2$ with $\tau_0=\hbar\sqrt{N}/(c_1 n)$.}
 \label{fig1}
\end{figure}

To confirm the above result, we perform numerical simulations.
Because the interactions between excited atoms are neglected in the above discussion, we consider only the $\kk=\bm 0$ modes and simulate the spin mixing dynamics at $q=0$ following the simplified Hamiltonian
\begin{align}
 \hat{H}_0 =- \frac{c_1}{2\Omega}
[(\dhaa{1})^2-(\dhaa{0})^2-(\dhaa{-1})^2][\haa{1}^2-\haa{0}^2-\haa{-1}^2],
\label{eq:Hsimple}
\end{align}
where we omit the subscripts identifying the momentum.
Here, $\hat{H}_0$ comes from the second term in the square brackets in Eq.~\eqref{eq:Vint2}, because the other term in Eq.~\eqref{eq:Vint2} is a constant for a fixed number state.
We expand the $N$-particle state in terms of the Fock state $|r,s\rangle\equiv [r!s!(N-r-s)!]^{-1/2}(\dhaa{1})^{r}(\dhaa{0})^{N-r-s}(\dhaa{-1})^{s}\vac$, and numerically solve the dynamics starting from $|N_{\rm ex},N_{\rm ex}\rangle$.
Figure~\ref{fig1} shows the time evolution of $N_{1}/N$ for $N=250$ and $N_{\rm ex}=0,1$ and 2.
The initial growth of $N_1/N$ agrees well with the prediction of Eq.~\eqref{eq:barN}, showing that the growth of the zero mode is indeed enhanced due to bosonic stimulation.

Finally, we discuss the case for a trapped system.
In the previous studies that discuss many-body spin dynamics in a micro condensate~\cite{Law1998,Koashi2000,Pu1999,Zhou2001,Cui2008,Barnett2010,*Barnett2011}, the confining potential is assumed to be much stronger than the spin-dependent interactions so that all spin components share the same spatial dependence; this is called the single-mode approximation (SMA).
Because the motional degrees of freedom are neglected in the SMA, the effective Hamiltonian is the same as Eq.~\eqref{eq:Hsimple} if one replace the volume $\Omega$ with the effective volume $\Omega_{\rm eff}\equiv [\int d\rr |\psi(\rr)|^4]^{-1}$, where $\psi(\rr)$ is the common wavefunction of all spin components and normalized as $\int d\rr |\psi(\rr)|^2=1$.
Actually, $\tau_0$ obtained above coincides with the time scale for fragmentation in a micro condensate~\cite{Law1998,Cui2008}.
Even when the SMA is not applicable, it can be shown that there exists a pair of Goldstone magnons at $q=0$, which are superpositions of $m_F=1$ and $-1$ states with the same wavefunction as the condensate.
Our result indicates that the number of Goldstone magnons increases following Eq.~\eqref{eq:Nmk0}, as long as the condensate depletion is small.
For example, when $N=10^6$ of spin-1 sodium atoms are confined in a spherical trap with a trap frequency of $300$ Hz, the Thomas-Fermi distribution leads to $\tau_0=1.7$~s, where we use $a_2-a_0=2.47a_\textrm{B}$ with $a_\textrm{B}$ being the Bohr magneton~\cite{Black2007}.
The growth time is further reduced when the growing modes are initially populated:
If $N_\textrm{ex}=5$, for example, the growth time becomes $0.52$~s, which is short enough to observe in experiments~\cite{Bookjans2011,Vinit2013}.
A possible experimental scheme is to prepare a condensate in the $m_F=0$ state at $q=q_\textrm{ini}$ and suddenly change $q$ to zero at $t=0$.
For $q_\textrm{ini}>0$, the lowest magnon modes have the energy $\sim\sqrt{q_\textrm{ini}(q_\textrm{ini}+2c_1n_\textrm{eff})}$, where $n_\textrm{eff}\equiv N/\Omega_\textrm{eff}$, and are thermally populated according to the Bose-Einstein distribution function.
Hence, for a fixed temperature, the Goldstone magnons grow faster as $q_\textrm{ini}$ becomes smaller.

\section{Conclusion}
In conclusion, by employing the number-conserving Bogoliubov theory, we have shown that the Goldstone magnons in a polar BEC are unstable and lead to fragmentation when the system is finite.
The time scale for the magnon growth diverges in the thermodynamic limit and the polar state becomes stable.
The growth is further enhanced when the corresponding modes are initially occupied by thermal atoms, which can be tuned by changing the initial value of the quadratic Zeeman energy.
In a similar manner, the Goldstone modes, except for the Goldstone phonon, are shown to be unstable in other spinor systems when the real ground state does not break the spin rotational symmetry.
The instability of the polar state at $q\le 0$ is experimentally investigated in Ref.~\cite{Bookjans2011}, where the decrease in the fraction of $m_F=0$ component is slower than the time predicted from the dynamical instability.
To understand this experiment, we may need to take into account the interactions between excitations and the subsequent thermalization of the system, which remains as a future study.

This work was supported by KAKENHI (Grants No. 22340114 and No. 22740265) from MEXT of Japan, the Funding Program for World-Leading Innovation R \& D on Science and Technology (FIRST), and the Inoue Foundation for Science.
YK acknowledges fruitiful discussions with M. Ueda.

\bibliography{reference}
\end{document}